\newif\ifmnras
\ifmnras

\documentclass[a4paper,fleqn,usenatbib]{mnras}
\usepackage{newtxtext,newtxmath}
\usepackage[T1]{fontenc}
\usepackage{ae,aecompl}

\usepackage{graphicx}	% Including figure files
\usepackage[section] {placeins}
\usepackage{subfigure}
\usepackage{float}
\usepackage{color}
\usepackage{hyperref}
\graphicspath{{./figures/}}
\usepackage{float}

\def\msun{{\rm\,M_\odot}} 
\def\lsun{{\rm\,L_\odot}}
\def\zsun{{\rm\,Z_\odot}}
\newcommand{\etal}{et al.\ }
\newcommand{\kms}{\, {\rm km\, s}^{-1}}
\newcommand{\ikms}{(\kms)^{-1}}
\newcommand{\mpc}{\, {\rm Mpc}}
\newcommand{\kpc}{\, {\rm kpc}}
\newcommand{\hmpc}{\, h^{-1} \mpc}
\newcommand{\ihmpc}{(\hmpc)^{-1}}
\newcommand{\hkpc}{\, h^{-1} \kpc}
\newcommand{\lya}{Ly$\alpha$}
\newcommand{\lyaf}{Ly$\alpha$ forest}
\newcommand{\ch}{\bf change}
\newcommand{\gmo}{{\gamma-1}}
\newcommand{\bF}{\bar{F}}
\newcommand{\hi}{\mbox{H\,{\scriptsize I}\ }}
\newcommand{\heii}{\mbox{He\,{\scriptsize II}\ }}
\newcommand{\civ}{\mbox{C\,{\scriptsize IV}\ }}
\newcommand{\kpa}{k_\parallel}
\newcommand{\vk}{{\mathbf k}}
\newcommand{\df}{\delta_F}
\newcommand{\sF}{{F_s}}
\newcommand{\sdelta}{{\delta_s}}
\newcommand{\seta}{{\eta_s}}
\newcommand{\dt}{\Delta \theta}
\newcommand{\dv}{\Delta v}
\newcommand{\pa}{\parallel}
\newcommand{\pe}{\perp}
\newcommand{\dz}{\Delta z}
\newcommand{\llya}{L$_{{\rm Ly}\alpha}$}
\newcommand{\lheii}{L$_{{\rm He II}}$}
\newcommand{\lciv}{L$_{{\rm C IV}}$}
\newcommand{\expZ}{$\langle Z \rangle$}
\newcommand{\expT}{$\langle T \rangle$}
\newcommand{\expD}{$\langle n_{{\rm H}} \rangle$}
\newcommand{\expF}{$\langle f_{{\rm HI}} \rangle$}
\def\h2{${\rm\,H_2}$}

\title{Upper Limit on Star Formation and Metal Enrichment in Minihalos}

\author[]{Renyue Cen
\\
Department of Astrophysical Sciences, Princeton University, Princeton, NJ 08544\\
} 
\pubyear{2015}

\begin{document}
\label{firstpage}
\pagerange{\pageref{firstpage}--\pageref{lastpage}}
\maketitle

%%%%%%%% MNRAS VERSION END %%%%%%%%
%%%%%%%%%%%%%%%%%%%%%%%%%%%%%%%%%%%%%%%%%%%%%%%%%
%%%%%%%%%%%%%%%%%%%%%%%%%%%%%%%%%%%%%%%%%%%%%%%%%
%%%%%%%%%%%%%%%%%%%%%%%%%%%%%%%%%%%%%%%%%%%%%%%%%
%%%%%%%%%%%%%%%%%%%%%%%%%%%%%%%%%%%%%%%%%%%%%%%%%
%%%%%%%%%%%%%%%%%%%%%%%%%%%%%%%%%%%%%%%%%%%%%%%%%
%%%%%%%%%%%%%%%%%%%%%%%%%%%%%%%%%%%%%%%%%%%%%%%%%
%%%%%%%%%%%%%%%%%%%%%%%%%%%%%%%%%%%%%%%%%%%%%%%%%
%%%%%%%%%%%%%%%%%%%%%%%%%%%%%%%%%%%%%%%%%%%%%%%%%
%%%%%%%%%%%%%%%%%%%%%%%%%%%%%%%%%%%%%%%%%%%%%%%%%
%%%%%%%%%%%%%%%%%%%%%%%%%%%%%%%%%%%%%%%%%%%%%%%%%
%%%%%%%% PREPRINT VERSION START %%%%%%%%
\else

\pdfoutput=1
\documentclass[12pt,preprint]{aastex}

\textheight=9.2in
\topmargin=-0.5in
\textwidth=6.5in
\rightmargin=2.0in

\usepackage[T1]{fontenc}
\usepackage{ae,aecompl}

\usepackage{graphicx}	% Including figure files
\usepackage[section] {placeins}
\usepackage{subfigure}
\usepackage{float}
\usepackage{color}
\usepackage{hyperref}
\graphicspath{{./figures/}}
\usepackage{float}

\usepackage[scaled]{helvet}
\renewcommand*\familydefault{\sfdefault}
\usepackage[T1]{fontenc}

\def\msun{{\rm\,M_\odot}} 
\def\lsun{{\rm\,L_\odot}}
\def\zsun{{\rm\,Z_\odot}}
\newcommand{\etal}{et al.\ }
\newcommand{\kms}{\, {\rm km\, s}^{-1}}
\newcommand{\ikms}{(\kms)^{-1}}
\newcommand{\mpc}{\, {\rm Mpc}}
\newcommand{\kpc}{\, {\rm kpc}}
\newcommand{\hmpc}{\, h^{-1} \mpc}
\newcommand{\ihmpc}{(\hmpc)^{-1}}
\newcommand{\hkpc}{\, h^{-1} \kpc}
\newcommand{\lya}{Ly$\alpha$}
\newcommand{\lyaf}{Ly$\alpha$ forest}
\newcommand{\ch}{\bf change}
\newcommand{\gmo}{{\gamma-1}}
\newcommand{\bF}{\bar{F}}
\newcommand{\hi}{\mbox{H\,{\scriptsize I}\ }}
\newcommand{\heii}{\mbox{He\,{\scriptsize II}\ }}
\newcommand{\civ}{\mbox{C\,{\scriptsize IV}\ }}
\newcommand{\kpa}{k_\parallel}
\newcommand{\vk}{{\mathbf k}}
\newcommand{\df}{\delta_F}
\newcommand{\sF}{{F_s}}
\newcommand{\sdelta}{{\delta_s}}
\newcommand{\seta}{{\eta_s}}
\newcommand{\dt}{\Delta \theta}
\newcommand{\dv}{\Delta v}
\newcommand{\pa}{\parallel}
\newcommand{\pe}{\perp}
\newcommand{\dz}{\Delta z}
\newcommand{\llya}{L$_{{\rm Ly}\alpha}$}
\newcommand{\lheii}{L$_{{\rm He II}}$}
\newcommand{\lciv}{L$_{{\rm C IV}}$}
\newcommand{\expZ}{$\langle Z \rangle$}
\newcommand{\expT}{$\langle T \rangle$}
\newcommand{\expD}{$\langle n_{{\rm H}} \rangle$}
\newcommand{\expF}{$\langle f_{{\rm HI}} \rangle$}
\def\h2{${\rm\,H_2}$}

\title{Upper Limit on Star Formation and Metal Enrichment in Minihalos}
\author{Renyue Cen$^{1}$} 

\begin{document}
\label{firstpage}

%%%%%%%% PREPRINT VERSION END %%%%%%%%
%%%%%%%%%%%%%%%%%%%%%%%%%%%%%%%%%%%%%%%%%%%%%%%%%
%%%%%%%%%%%%%%%%%%%%%%%%%%%%%%%%%%%%%%%%%%%%%%%%%
%%%%%%%%%%%%%%%%%%%%%%%%%%%%%%%%%%%%%%%%%%%%%%%%%
%%%%%%%%%%%%%%%%%%%%%%%%%%%%%%%%%%%%%%%%%%%%%%%%%
%%%%%%%%%%%%%%%%%%%%%%%%%%%%%%%%%%%%%%%%%%%%%%%%%
%%%%%%%%%%%%%%%%%%%%%%%%%%%%%%%%%%%%%%%%%%%%%%%%%
%%%%%%%%%%%%%%%%%%%%%%%%%%%%%%%%%%%%%%%%%%%%%%%%%
%%%%%%%%%%%%%%%%%%%%%%%%%%%%%%%%%%%%%%%%%%%%%%%%%
%%%%%%%%%%%%%%%%%%%%%%%%%%%%%%%%%%%%%%%%%%%%%%%%%
%%%%%%%%%%%%%%%%%%%%%%%%%%%%%%%%%%%%%%%%%%%%%%%%%
\fi

\begin{abstract}

An analysis of negative radiative feedback from resident stars in minihalos is performed.
It is found that the most effective mechanism to suppress star formation is 
provided by infrared photons from resident stars via  
photo-detachment of ${\rm H^-}$. 
It is shown that a stringent upper bound on (total stellar mass, metallicity) of 
($\sim 1000\msun$, $-3.3\pm 0.2$) in any newly minted atomic cooling halo can be placed, 
with the actual values possibly significantly lower. 
This has both important physical ramifications on formation of stars and 
supermassive black seeds in atomic cooling halos at high redshift,
pertaining to processes of low temperature metal cooling, dust formation and fragmentation,
and direct consequences on the faint end galaxy luminosity function at high redshift 
and cosmological reionization.
The luminosity function of galaxies at the epoch of reionization may be substantially affected due to the combined effect
of a diminished role of minihalos and an enhanced contribution from Pop III stars in atomic cooling halos.
Upcoming results on reionization optical depth from Planck High-Frequency Instrument
data may provide a significant constraint on and a unique probe of 
this star formation physical process in minihalos.
As a numerical example, in the absence of significant contributions from minihalos
with virial masses below $1.5\times 10^{8}\msun$
the reionization optical depth is expected to be no greater than $0.065$,
whereas allowing for minihalos of masses as low as
($10^7\msun$, $10^{6.5}\msun$) to form stars unconstrained by this self-regulation physical process,
the reionization optical depth is expected to exceed $(0.075,0.085)$, respectively.

\end{abstract}

\ifmnras

\begin{keywords}
stars: formation,
stars: Population III,
ISM: abundances,
galaxies: formation,
galaxies: ISM
\end{keywords}

else

%\keywords{
%stars: formation -
%stars: Population III -
%ISM: abundances -
%galaxies: formation -
%galaxies: ISM
%}

\fi
% the above seems to cause latex not to work for MNRAS version

\section{Introduction}

Star formation in minihalos is a fundamental issue,
because it is responsible for enriching the primordial gas with first metals that 
shape the subsequent formation of stars and possibly supermassive black hole seeds in atomic cooling halos.
Since the pioneering works \citep[e.g.,][]{2002Abel, 2002Bromm, 2002Nakamura},
most studies have focused on formation of individual stars \citep[e.g.,][]{2014Hirano}.
So far studies of the effects of external 
Lyman-Werner band (LW) ($h\nu=11.2-13.6$eV) radiation background \citep[e.g.,][]{2001Machacek, 2007Wise, 2008OShea},
external IR radiation background \citep[e.g.,][]{2007Chuzhoy,2015Hirano}
on gas chemistry and thermodynamics hence star formation in minihalos have produced significant physical insight.
We assess the effects of these two - LW photo-dissociation and IR photo-detachment - \h2 formation suppressing processes 
due to resident stellar population within minihalos, instead of the respective collective backgrounds widely considered.
%Here we examine a self-regulating effect due to star formation in minihalos:
%how internal radiative feedback from resident stars in minihalos regulates/suppresses
%\h2 formation hence subsequent star formation.
We show that photo-detachment process of \h2 
by infrared photons of energy $h\nu\ge 0.755$eV produced by resident stars 
places a strong upper bound on stellar mass and metals that may be formed in minihalos.
This upper limit needs to be taken into account in the general considerations of galaxy formation 
at high redshift.

\section{Maximum Stellar Mass and Metal Enrichment in Minihalos}

Minihalos are defined as small dark matter halos with virial temperature below that for efficient atomic cooling
(i.e., ${\rm T_v\le 10^4}$K). 
Minihalos form early in the standard cold dark matter model and are only relevant for high redshift.
Star formation may start in minihalos with ${\rm T_v}$ as low as $\sim 1000$K or so.
The relation between halo virial mass (${\rm M_v}$) and virial temperature (${\rm T_v}$) is 
\begin{equation}
M_{\rm v} = 10^8h^{-1}\msun \left(\frac{T_{\rm
v}}{1.98\times10^{4}K}\right)^{\frac{3}{2}}\left(\frac{0.6}{\mu_P}\right)^\frac{3}{2}
\left(\frac{\Omega_{\rm
m}}{\Omega_m^z}\frac{\Delta_c}{18\pi^2}\right)^{-\frac{1}{2}}\left(\frac{1+z}{10}\right)^{-\frac{3}{2}},
\label{eq:Mv}
\end{equation}
where $z$ is redshift,
$\Omega_m$ and $\Omega_\Lambda$
are density parameter and cosmological constant at redshift zero, respectively;
$\Omega_m^z\equiv[1+(\Omega_\Lambda/\Omega_m)(1+z)^{-3}]^{-1}$ is the density parameter at redshift $z$;
$\Delta_c=18\pi^2+82d-39d^2$ and $d=\Omega_m^z-1$ (see Barkana \& Loeb~2001 for more details).
The corresponding physical virial radius is 
\begin{equation}
r_{\rm v} = 0.784{\rm h^{-1}kpc} \left(\frac{T_{\rm
v}}{1.98\times10^{4}K}\right)^{\frac{1}{2}}\left(\frac{0.6}{\mu_P}\right)^\frac{1}{2}
\left(\frac{\Omega_{\rm
m}}{\Omega_m^z}\frac{\Delta_c}{18\pi^2}\right)^{-\frac{1}{2}}\left(\frac{1+z}{10}\right)^{-\frac{1}{2}}.
\label{eq:rv}
\end{equation}

\noindent
In minihalos at high redshift, molecular hydrogen \h2 is the primary gas cooling agent, before a significant amount of metals is present.
In the absence of a significant amount of dust grains,
the dominant \h2 formation channel is via a two-step gas phase process \citep[e.g.,][]{2003Draine},
first with radiative association:
\begin{equation}
\label{eq:ra}
{\rm H} + e^{-} \rightarrow {\rm H}^- + h\nu,
\end{equation}
\noindent
followed by associative detachment:
\begin{equation}
\label{eq:sd}
{\rm H^{-}} + {\rm H} \rightarrow {\rm H_2} + e^-.
\end{equation}
\noindent
Given this formation channel, if one is interested in suppressing \h2 formation,
there are two main ways to achieve that goal.
One is by destruction of formed \h2 molecules through the photo-dissociation process by photons 
in the LW band of $h\nu=11.2-13.6$eV:
\begin{equation}
\label{eq:pd}
{\rm H^2} + h\nu \rightarrow {\rm H} + {\rm H}.
\end{equation}
\noindent
The other is by reducing the density of ${\rm H^{-}}$, to which the rate of \h2 formation is proportional,
by infrared (IR) photons of energy $h\nu\ge 0.755$eV via the photo-detachment process:
\begin{equation}
\label{eq:pde}
{\rm H^-} + h\nu \rightarrow {\rm H} + e^-.
\end{equation}
\noindent

For simplicity, we assume that the initial mass function (IMF) of Population III (Pop III) stars has
a powerlaw distribution of the same Salpeter slope:
\begin{equation}
\label{eq:imf}
{\rm n(M_*)dM_* = C M^{-2.35}dM_*},
\end{equation}
\noindent
with an upper mass cutoff $100\msun$ and 
a lower mass cutoff ${\rm M_{low}}$ that we will vary to understand its influence on the results;
${\rm C}$ is a constant normalizing the stellar abundance per unit of star formation rate.
We stress that our results are rather insensitive to either ${\rm M_{low}}$ or the slope of the IMF.
Then, one can compute the intrinsic spectral luminosity (in units of ${\rm erg~sec^{-1}~Hz^{-1}~sr^{-1}}$)
per stellar mass at any photon energy $\nu$ as
\begin{equation}
\label{eq:LX}
{
\rm L_{\nu} = \int_{0}^{t_h} {\int_{L_{low}}^{100\msun} {\theta}(t_{ms}-t_h+t_f) \dot M_*(t_f)J_\nu(M_*) n(M_*)dM_*dt_f}
},
\end{equation}
\noindent
where ${\rm J_\nu(M_*)}$ is the mean spectral luminosity of a star of mass ${\rm M_*}$ at photon energy $h\nu$ in the main sequence;
$\theta(x)$ is the Heaviside theta function;
${\rm t_{ms}(M_*)}$ is the star's main sequence lifetime;
${\rm t_f}$ and ${\rm t_h}$ are the formation time of the star in question
and the time under consideration when the luminosity is computed;
${\rm \dot M_*(t_f)}$ is star formation rate at time ${\rm t_f}$.

\ifmnras

\begin{figure}
\centering
\includegraphics[width=3.5in, keepaspectratio]{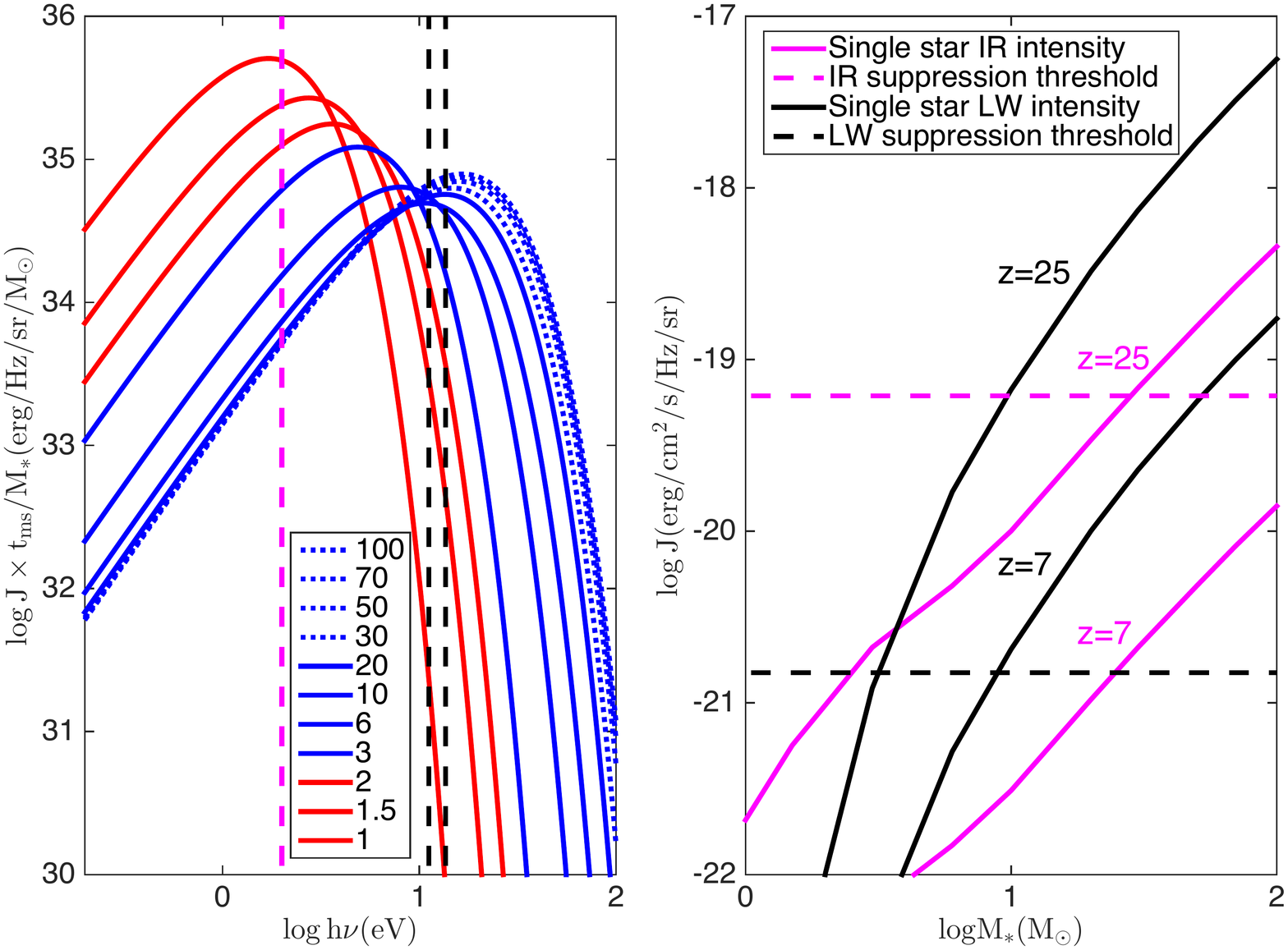}
\caption{
{\color{red}Left panel:}
shows the intrinsic 
black-body spectra of individual Pop-III stars per unit stellar mass,
multiplied by the main sequence lifetime, for a set of stellar masses (in units of solar mass) indicated in the legend. 
Also shown as the vertical magenta dashed line is the photon energy of $2$eV for the photo-detachment process.
The LW band is indicated by the two black vertical dashed lines.
{\color{red}Right panel:}
shows the radiation intensity at $2$eV (magenta solid curves) and $11.2$eV (black solid curves)
for a single star of mass shown on the x-axis.
The star is assumed to be located at the center and the intensity is measured 
at the core radius of the minihalo (see text for definition of core radius).
Two cases are shown, one for a minihalo at $z=25$ with virial temperature of $10^3$~K (upper solid curves)
and the other at $z=7$ with virial temperature of $10^4$~K (lower solid curves).
For both IR and LW photons, no absorption is assumed for this illustration.
The horizontal dashed lines with the same corresponding colors are the threshold
intensity for complete suppression of \h2 formation by the respective processes.
}
\label{fig:bb}
\end{figure}

\else

\begin{figure}[!h]
\begin{center}
\includegraphics[width=6.5in, keepaspectratio]{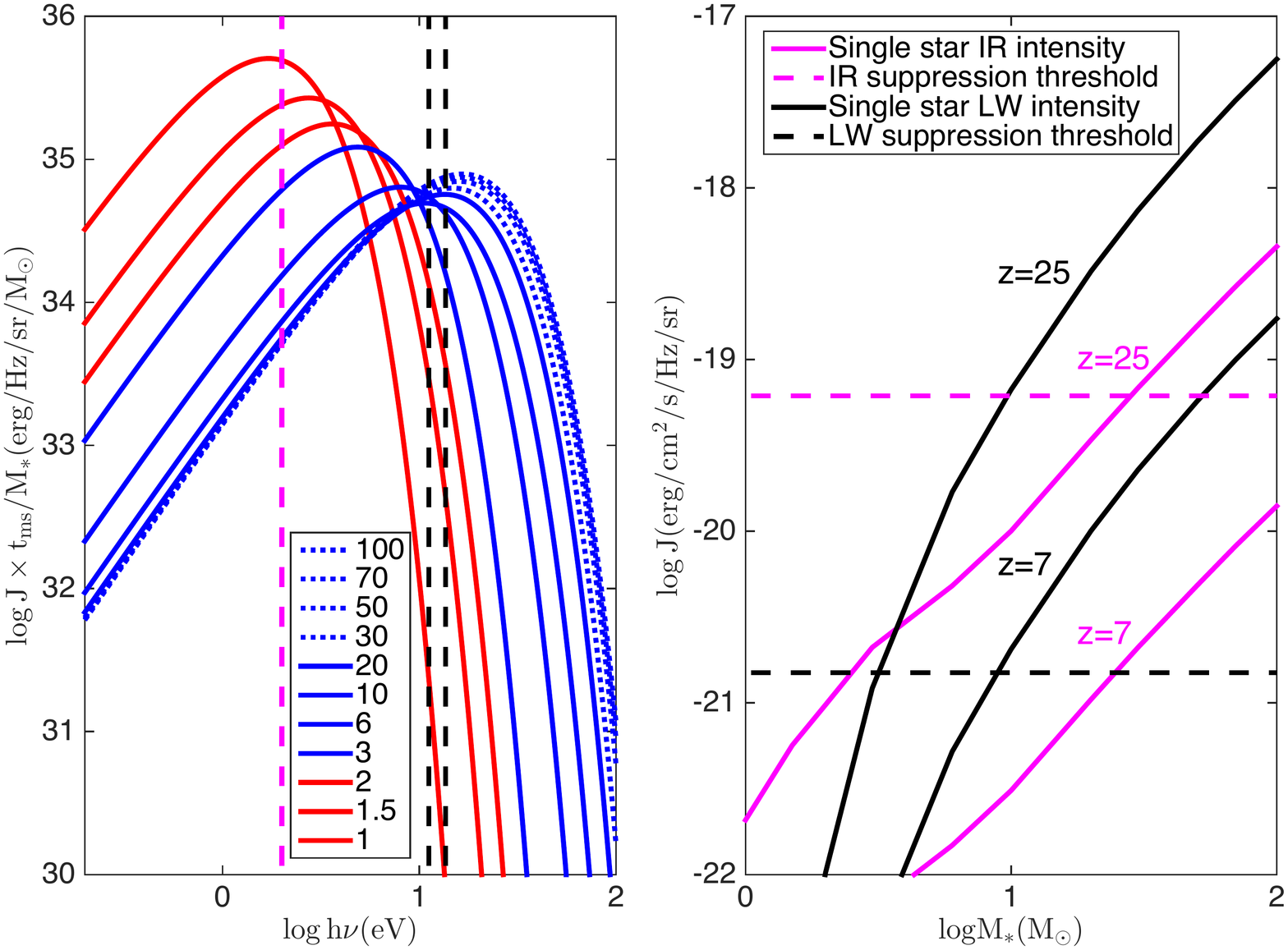}
\end{center}
\vskip -1.0cm
\caption{
{\color{red}Left panel:}
shows the intrinsic 
black-body spectra of individual Pop-III stars per unit stellar mass,
multiplied by the main sequence lifetime, for a set of stellar masses (in units of solar mass) indicated in the legend. 
Also shown as the vertical magenta dashed line is the photon energy of $2$eV for the photo-detachment process.
The LW band is indicated by the two black vertical dashed lines.
{\color{red}Right panel:}
shows the radiation intensity at $2$eV (magenta solid curves) and $11.2$eV (black solid curves)
for a single star of mass shown on the x-axis.
The star is assumed to be located at the center and the intensity is measured 
at the core radius of the minihalo (see text for definition of core radius).
Two cases are shown, one for a minihalo at $z=25$ with virial temperature of $10^3$~K (upper solid curves)
and the other at $z=7$ with virial temperature of $10^4$~K (lower solid curves).
For both IR and LW photons, no absorption is assumed for this illustration.
The horizontal dashed lines with the same corresponding colors are the threshold
intensity for complete suppression of \h2 formation by the respective processes.
}
\label{fig:bb}
\end{figure}

\fi

The left panel of Figure~\ref{fig:bb} shows 
the individual intrinsic Pop-III stellar black-body spectrum per unit stellar mass 
times the main sequence lifetime 
for a range of masses for individual Pop III stars (indicated in the legend in units of solar mass), 
based on data from \citet[][]{2001Marigo}.
It is easy to see that low mass stars are more efficient producers of IR photons 
(indicated by the vertical dashed magenta line); for $1\msun$ to $20\msun$, a decrease of approximately 100 
for IR intensity per unit stellar mass is observed.
For the LW band photons
(indicated by the two vertical dashed black lines),
the opposite holds: a decrease of approximately four orders of magnitude 
is seen from $20\msun$ to $1\msun$.
In the right panel of Figure~\ref{fig:bb} we show
comparisons the IR (LW) intensities of a single star of mass indicated by the x-axis
in magenta (black) solid curves for redshifts $z=25$ ($z=7$), to be compared to the 
the threshold intensities for completion suppression of \h2 formation by the respective processes
shown as the horizontal dashed lines with the corresponding colors.
See below for how the the threshold intensities are computed.

\citet[][]{2012WolcottGreen} show that complete suppression of \h2 formation in minihalos at high redshift 
is possible by either LW photo-dissociation or IR photo-detachment process.
Based on a detailed modeling,
they derive a critical radiation intensity for complete suppression of \h2 formation of 
\begin{equation}
\label{eq:lw}
{\rm J_{LW,crit}=1.5\times 10^{-21}erg s^{-1} cm^{-2} Hz^{-1} sr^{-1}} 
\end{equation}
\noindent
at the LW band via photo-dissociation process alone,
and a critical radiation intensity of 
\begin{equation}
\label{eq:ir}
{\rm J_{IR,crit}=6.1\times 10^{-20}erg s^{-1} cm^{-2} Hz^{-1} sr^{-1}} 
\end{equation}
\noindent
at the IR band ($h\nu=2$eV) via photodetachment process alone,
under the assumption of the existence of the respective backgrounds, {\it not} internal radiation.

We consider the requirement of suppression 
of either \h2 or ${\rm H^{-}}$ formation in the central core region of minihalos,
which is likely most stringent compared to less dense gas at larger radii.
Following \citet[][]{1999Shapiro} we adopt
the core radius and density to be ${\rm r_{c}=r_v/29.4}$, which is then
\begin{equation}
{\rm r_{\rm c} = 26.7{\rm h^{-1}pc} \left(\frac{T_{\rm
v}}{1.98\times10^{4}K}\right)^{\frac{1}{2}}\left(\frac{0.6}{\mu_P}\right)^\frac{1}{2}
\left(\frac{\Omega_{\rm
m}}{\Omega_m^z}\frac{\Delta_c}{18\pi^2}\right)^{-\frac{1}{2}}\left(\frac{1+z}{10}\right)^{-\frac{1}{2}}
}.
\label{eq:rc}
\end{equation}
and 
hydrogen number density in the core is ${\rm n_{c}=514n_v}$ (${\rm n_v}$ is the gas number density at the virial radius):
\begin{equation}
{\rm n_{\rm c} = 4.5{\rm cm^{-3}} 
\left(\frac{1+z}{10}\right)^{3}
}.
\label{eq:rc}
\end{equation}
\noindent
Since we use the numerical results from \citet[][]{2012WolcottGreen} on photo-detachment,
it will be instructive to gain a physical understanding of its origin. 
%We demonstrate now how we can understand the results obtained by \citet[][]{2012WolcottGreen} on photo-detachment as follows.
The photodetachment cross section is 
\begin{eqnarray}
\label{eq:Fs}
{\rm \sigma_- = 2.1\times 10^{-16}{(\epsilon-0.755)^{3/2}\over \epsilon^{3.11}} cm^2}
\end{eqnarray}
\noindent
where $\epsilon$ is the photon energy in units of eV.
%giving ${\rm \sigma_-(2eV) = 3.4\times 10^{-17}~cm^2}$. 
The radiative association rate coefficient is $k_-=1.3\times 10^{-9}$cm$^3$~s$^{-1}$.
Thus, with ${\rm J_{IR,crit}=6.1\times 10^{-20}erg/s/cm^2/Hz/sr}$ at $2$eV
and minihalo core density of ${\rm n_{c}=31cm^{-3}}$ at $z=18$ (see Equation \ref{eq:Mv})
($z=18$ is used in \citet[][]{2012WolcottGreen}) and assuming that the spectrum shape of $\propto \nu^{0}$ in the range $0.755-13.6$eV, one finds that
the ratio of photo-detachment rate to radiative association rate is 0.46;
the ratio becomes $1.5$ if one assumes the spectrum shape of $\propto \nu^{+1}$.
Note that in the Raleigh-Jeans limit the spectral shape goes as $\propto \nu^{+2}$ (see Figure~\ref{fig:bb}).
We now see that when the photo-detachment rate and radiative association rate 
are approximately equal in the minihalo core, \h2 formation is effectively completely suppressed,
as one would have expected.
This thus provides an order of magnitude understanding of the \citet[][]{2012WolcottGreen} results.

Given the expected little dust content in very metal poor gas in minihalos, 
the optical depth for IR photons at $2$eV is negligible.
As a numerical example, the core hydrogen column density 
would be ${\rm N_c\equiv r_{c}n_{c}=1.5\times 10^{20}cm^{-2}}$
for a minihalo of ${\rm T_v=10^4}~$K at $z=8$.
Using the gas to dust column ratio \citep[][]{2003Draine} with the assumption
that dust content is linearly proportional to metallicity yields
$A_V=0.08(Z/\zsun)$ mag in this case.
It is easy to see that we may safely neglect optical depth effect for IR photons in question.
For LW photons, \h2 self-shielding effect may be important.
We include, conservatively, for maximum \h2 self-shielding of LW radiation by placing all sources
at the center of the minihalo with the self-shielding reduction of LW photons 
using the accurate fitting formula
from \citet[][]{1996Draine} for a halo at $z=7$ with $T_v=10^4$~K, 
corresponding Doppler parameter $b=13\kms$, 
\h2 fraction of ${\rm f_{H_2}=10^{-3}}$ and \h2 column density equal to ${\rm f_{H_2} r_{c}n_{c}}$.
This case is contrasted with the hypothetical case where self-shielding is neglected.

\ifmnras

\begin{figure}
\centering
\includegraphics[width=3.5in, keepaspectratio]{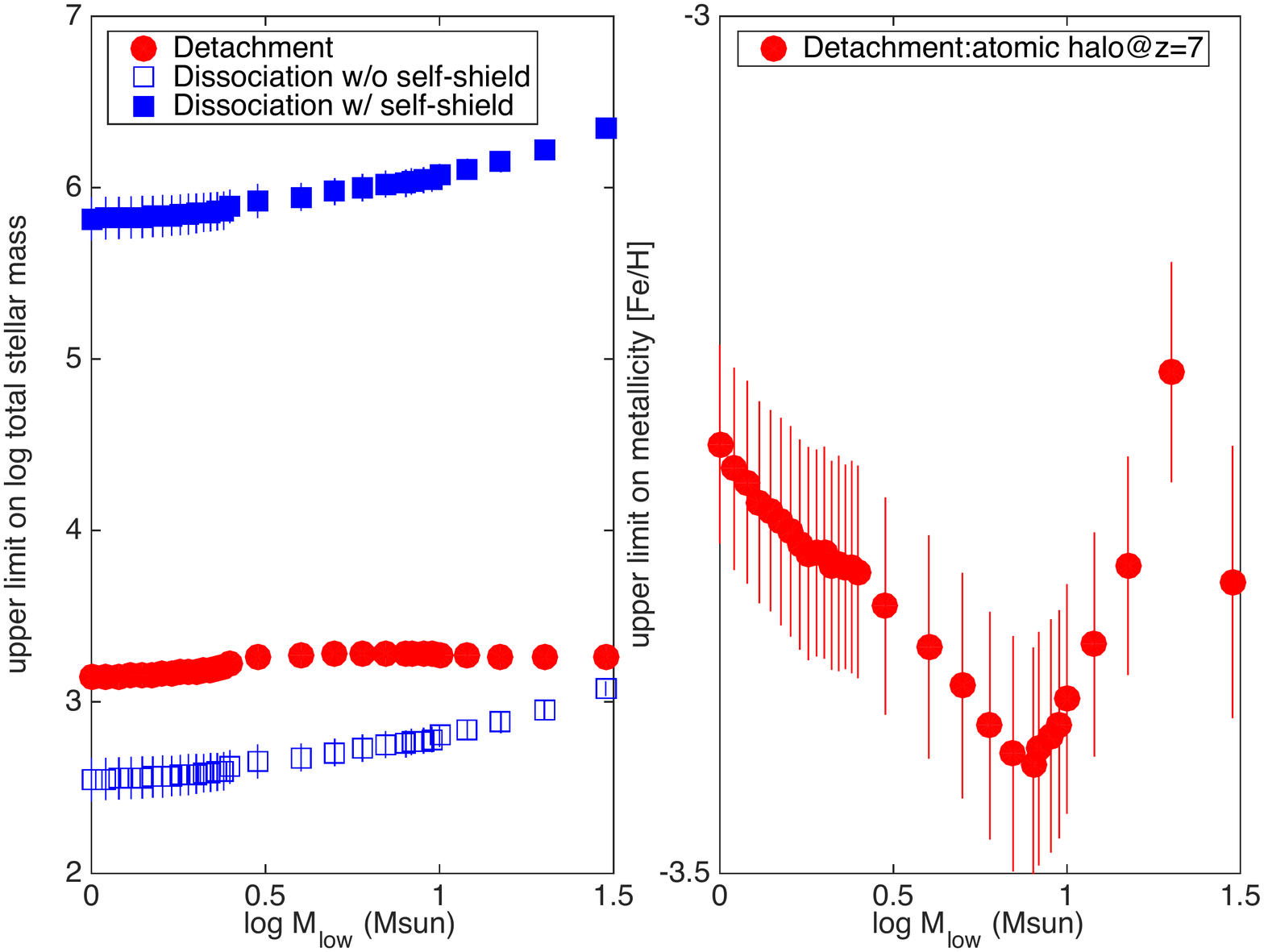}
\caption{
{\color{red}Left panel:}
shows the critical cumulative stellar mass for complete suppression of \h2 formation,
as a function of the lower mass cutoff of the IMF ${\rm M_{low}}$,
via either 
the photo-dissociation process by LW photons 
with (blue open squares) and without \h2 self-shielding of LW photons (blue solid squares)
or 
the photo-detachment process by infrared photons (red solid dots).
In this example, we assume that a minihalo of virial temperature $T_v=10^3$K is formed at $z=25$ when
star formation commences, and the critical stellar mass (i.e., upper limit on total stellar mass) 
is evaluated at $z=7$ when the minihalo has grown to a virial temperature of $T_v=10^4$K.
{\color{red}Right panel:}
shows the upper bound on the mean gas metallicity,
corresponding to the critical stellar mass 
shown in the left panel,
evaluated at $z=7$ when the minihalo has grown to a virial temperature of $T_v=10^4$K.
In both panels, the errorbars indicate the dispersions obtained by Monte Carlo
realizations of different star formation histories, described in the text.
}
\label{fig:Z}
\end{figure}

\else

\begin{figure}[!h]
\begin{center}
\vskip -0.5cm
\includegraphics[width=6.5in, keepaspectratio]{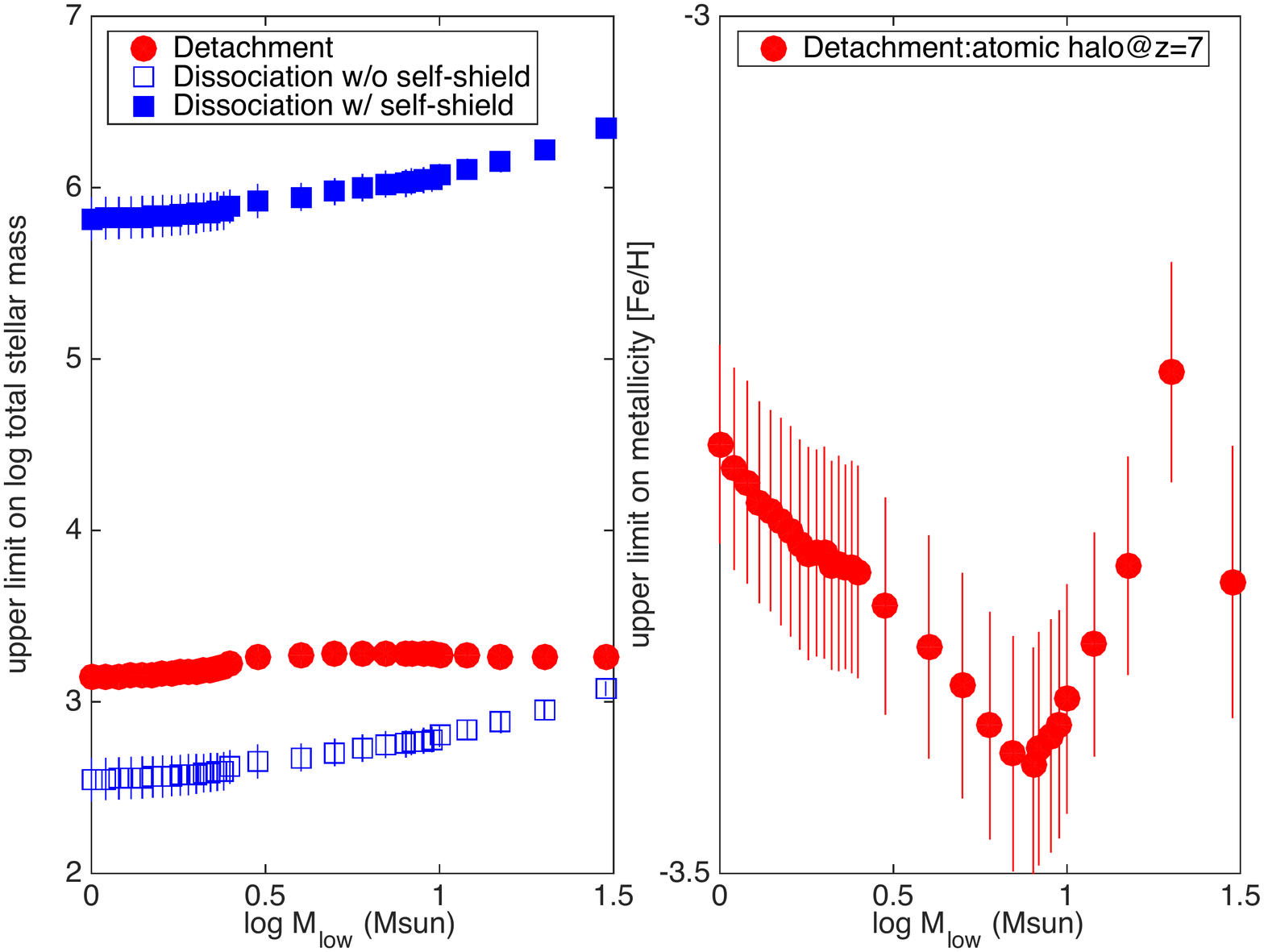}
\hskip -1cm
\end{center}
\vskip -0.5cm
\caption{
{\color{red}Left panel:}
shows the critical cumulative stellar mass for complete suppression of \h2 formation,
as a function of the lower mass cutoff of the IMF ${\rm M_{low}}$,
via either 
the photo-dissociation process by LW photons 
with (blue open squares) and without \h2 self-shielding of LW photons (blue solid squares)
or 
the photo-detachment process by infrared photons (red solid dots).
In this example, we assume that a minihalo of virial temperature $T_v=10^3$K is formed at $z=25$ when
star formation commences, and the critical stellar mass (i.e., upper limit on total stellar mass) 
is evaluated at $z=7$ when the minihalo has grown to a virial temperature of $T_v=10^4$K.
{\color{red}Right panel:}
shows the upper bound on the mean gas metallicity,
corresponding to the critical stellar mass 
shown in the left panel,
evaluated at $z=7$ when the minihalo has grown to a virial temperature of $T_v=10^4$K.
In both panels, the errorbars indicate the dispersions obtained by Monte Carlo
realizations of different star formation histories, described in the text.
}
\label{fig:Z}
\end{figure}

\fi

The left panel of Figure~\ref{fig:Z} shows the critical cumulative stellar mass 
required to completely suppress further star formation,
as a function of the lower mass cutoff of the IMF ${\rm M_{low}}$,
following a minihalo of ${\rm T_v=10^3}$K at $z=25$ through its becoming an atomic cooling halo at $z=7$.
In making this plot, we have adopted a Monte Carlo approach to randomly sample the IMF,
assuming each starburst lasts about $4$Myr, a time scale to approximate the effect of supernova blowout.
While simulations have shown that the separation of episodic starbursts 
is about $20-100$Myr \citep[e.g.,][]{2014Kimm} for atomic cooling halos,
we expect that the separations for minihalos would be larger, thanks to 
the more violent blowouts of gas by supernovae out of shallower potential wells and 
less efficient cooling in minihalos for gas return.
To stay on the conservative side, we use temporal separations between star formation episodes of $20$Myr.
In general, a larger separation gives a lower total stellar mass, because 
the radiative suppression effects are almost entirely dominated by stars formed within the ongoing starburst
(not by stars from previous starbursts) and often the radiation from a 
single star is enough to provide the necessary suppression (see the right panel of Figure~\ref{fig:bb}).
On details regarding the Monte Carlo realizations, 
within each starburst, we randomly draw stars from the IMF with a lower mass cutoff of ${\rm M_{low}}$,
until the radiation intensity in IR or UV, separately,
at the core radius exceeds the required threshold.
We keep track of stars formed in starbursts at higher redshift and 
take into account their radiative contributions given their main sequence lifetimes.
Since we can not ``draw" a fractional star, 
in cases where a single star would already exceed the required threshold,
stellar mass is higher than if fractional stars can be drawn.
Based on the Monte Carlo random sampling procedure to draw stellar distribution from the IMF, 
the obtained dispersion are shown as vertical bars on symbols in 
both panels of Figure~\ref{fig:Z}.

It is evident that, taking into account \h2 self-shielding of LW photons,
for the entire range of ${\rm M_{low}}$ considered,
the destructive effect due to photo-detachment is larger by two-three 
orders of magnitude than that due to photo-dissociation taking into account attenuation for LW photons.
Thus, we will use the photo-detachment effect to place an upper bound on stellar mass that can form
before further \h2 formation hence star formation is completely suppressed.
The amount of stars formed within minihalos is small, at $\sim 10^3\msun$, prior to the minihalo becoming an atomic halo. 
This self-regulation of star formation in minihalos likely have a
significant impact on the possible contribution of minihalos to reionization.
A full characterization of this effect would need detailed simulations with this important process included.
\citet[][]{2014Wise} find stellar mass of $10^{3.5}-10^{4.0}\msun$
in minihalos of mass $10^{6.5}-10^{7.5}\msun$,
which is approximately a factor of at least $3-10$ higher than allowed,
even compared to the largest possible minihalos (before their becoming atomic cooling halos) considered here,
as shown in the left panel of Figure~\ref{fig:Z}.
We note that the amount stars formed are a result of accumulation of the number of star formation episodes.
We have "maximized" the stellar mass by using a conservative episodic interval and considering
the maximum minihalos at a low redshift $z=7$.
Obviously, for smaller minihalos at higher redshift with longer ``quiet" periods 
the amount of stellar mass formed will be smaller.
This suggests that the contribution of stars formed in minihalos to reionization may be substantially reduced.
We estimate that the contribution of minihalos to
cosmological reionization photon budget is likely limited to a few percent.

Next, we consider the metal enrichment due to stars formed in minihalos.
To compute that, we use the relation between the nickel (which decays to iron)
mass produced by a supernova of mechanical explosion energy $E$:
\begin{equation}
{\rm \log {M_{ni}\over \msun} = 1.49\log {E\over 10^{50}~erg} - 2.9}
\label{eq:Mni}
\end{equation}
\noindent
\citep[][]{2015Pejcha} and the relation between explosion energy $E$ and the main sequence stellar mass $M$:
\begin{equation}
{\rm {E\over 10^{51}~erg} = ({M\over 10.8\msun})^2}
\label{eq:Mni}
\end{equation}
\noindent
\citep[][]{2013Poznanski}.
We assume all stars with main sequence mass above $8\msun$ explode as supernavae,
except the two intervals $17-23\msun$ and $\ge 40\msun$, which produce black holes
based on the so-called compact parameter $\xi$ as a physical variable \citep[e.g.,][]{2011OConnor, 2015bPejcha}.

The right panel of Figure~\ref{fig:Z} shows the expected average metallicity 
when an atomic cooling halo is reached at $z=7$.
corresponding to the critical stellar mass shown as solid red dots in the left panel of 
Figure~\ref{fig:Z}.
We see that, on average, the expected maximum metallicity due to stars formed in minihalos
falls into the range of $-3.3\pm 0.2$ in solar units, for ${\rm M_{low}=1-30\msun}$.
We use iron mass fraction of $1.77\times 10^{-3}$ as solar abundance \citep[][]{2009Asplund}.
We have conservatively assumed that enrichment process takes places in a closed-box fashion,
with respect to metals produced.
Furthermore, we have simplistically assumed that none of the metals produced is not incorporated back into subsequent stars.
In reality, retainment of metals produced by stars in minihalos 
is probably far from complete, given their shallow potential wells,
i.e., it is not a closed box.
Furthermore, some of the earlier produced metals inevitably get reformed into stars.
These conservative approaches used, along with 
our conservative adoption of $20$Myr starburst separation,
indicate that 
that the actual metallicity
due to stars in minihalos may be significantly below the maximum allowed
values indicated in the right panel of Figure~\ref{fig:Z}.
In other words, we expect that the metallicity floor put in by 
stars formed in previous minihalos, when an atomic cooling halo is formed,
is likely significantly below $-3.3\pm 0.2$ in solar units.

There is one possible caveat in the arguments leading to the results.
Despite the resultant low metallicity due to self-suppression of star formation by
negative IR radiation feedback,
the metallicity is not zero.
Thus, it is prudent to check if the metallicity is sufficiently low to
justify the neglect of low-temperature metal cooling.
We find that, using ${\rm [Z/H]=-3}$ and molecular hydrogen fraction of ${\rm f_{H2}=10^{-3}}$, 
the ratio of the cooling rate of metal lines (primarily due to OI, CII, SiII aand FeII) to that of molecular hydrogen 
is found to be ($4.1\times 10^{-2}$, $1.6\times 10^{-3}$, $2.6\times 10^{-4}$) at 
temperature ${\rm T=(10^3, 10^{3.5}, 10^4)}~$K \citep[][]{2007Maio},
respectively.
Empirically, experimental simulations have found that, in lieu of molecular hydrogen cooling,
low-temperature metal cooling with a metallicity of ${\rm [Z/H]\sim -1.5}$ produces
cooling effect comparable to that molecular hydrogen fraction with ${\rm f_{H2}=10^{-3}}$
(Kimm 2016, private communications),
which is consistent with above estimates based on cooling rates.
Thus, the low-temperature metal cooling is probably no more than 
$~(\sim 2\%$, $0.1\%$, $0.01\%$) of the 
molecular hydrogen cooling in the case of absent negative feedback examined here, if ${\rm [Z/H]\le -3.3}$, 
in minihalos with virial temperatures ${\rm T_v=(10^3, 10^{3.5}, 10^4)}~$K, respectively.
Therefore, the low-temperature metal cooling is unlikely to be able to make up the ``lost" \h2 cooling,
due to negative feedback from local radiation,
to alter the suppression of star formation.

\section{Discussion and Conclusions}

This study investigates the radiative feedback from resident stars in minihalos.
We find that photo-detachment of ${\rm H^-}$ by infrared photons of energy $h\nu\ge 0.755$eV 
emitted by resident stars in minihalos is the most effective mechanism to 
suppress and hence self-regulate star formation within.
The negative feedback effect due to Lyman-Werner photons would have been more effective,
if the gas is transparent; however, \h2 self-shielding substantially reduces its effect to become
subdominant to that of photo-detachment process.

We find that the amount of stars formed in minihalos 
is capped at about $10^3\msun$, regardless of the lower mass cutoff of the initial mass function.
As a result, it is shown that a stringent upper bound of metallicity of $-3.3\pm 0.2$ 
relative to the solar value due to stars formed in minihalos can be placed; 
the actual amount of stars and metallicity achieved by stars in minihalos may be significantly lower,
because the various assumptions adopted, when needed, 
have been chosen to err, generously, on the conservative side to ensure
that our results with respect to star formation in minihalos represent an upper bound.

The self-regulation of star formation in minihalos likely has a
significant impact on the possible contribution of minihalos to reionization.
In \citet[][Figure 14]{2014Kimm} it is shown that,
in the absence of significant contributions from minihalos
with virial masses below $1.5\times 10^{8}\msun$, as an example, corresponding to minihalo threshold 
at $z=9$ (see Equation \ref{eq:rv}),
the reionization optical depth is expected to be no greater than $0.065$.
On the other hand, allowing for minihalos of masses as low as 
($10^7\msun$, $10^{6.5}\msun$) to form stars unconstrained by this self-regulation physical process,
the reionization optical depth would exceed $(0.075,0.085)$, respectively, 
in general agreement with earlier results under similar assumptions with respect 
to dramatically increased contributions especially with very massive Pop III stars \citep[e.g.,][]{2003Cen, 2003bCen, 2007Wyithe}.
While these values are all consistent with the most recent Planck results \citep[][$\tau_e=0.066\pm 0.016$]{2015Planck}
at $<1.2\sigma$ level,
upcoming results from Planck High-Frequency Instrument (HFI) 
data may provide a significant constraint on the star formation physics in minihalos.

The findings will also have profound ramifications on star formation and formation
of supermassive black seeds in atomic cooling halos at high redshift,
due to processes related to metal cooling, dust formation and fragmentation.
As an example, low-temperature metal cooling may be suppressed \citep[e.g.,][]{2003bBromm}
to increase the probability of extending the formation of Pop III stars.
Although simulations will be needed, 
this does suggest that, with a much lower mean metallicity, in conjunction with inhomogeneous metal enrichment processes,
pockets of Pop III stars in atomic cooling halos may be more widespread than thought. 
The combination of a reduction of star formation in minihalo 
and a possible increase in stellar luminosity in atomic cooling halos (due to Pop III stars) 
will alter both the slope and cutoff 
of the luminosity function of galaxies at the faint end at the epoch of reionization
\citep[][]{2014Kimm, 2015Trac}.
There may be two possible signatures in the luminosity function at the epoch of reionization.
First, a possible steepening at the faint end right before a dramatic flattening or downturn 
at the transition between the atomic cooling halo to minihalo mass may be expected;
the steepening is due to the increased proportion of metal-free stars in lower mass atomic cooling halos.
Second, due to generally increased variations in Pop III star fractions, in conjunction with stochastic starbursts,
the shape of the luminosity function at the high end is likely to resemble powerlaws than exponential.

On a separate, but potentially related subject,
we note that metallicities of stars in both types of globular clusters,
in the bimodal metallicity distribution \citep[e.g.,][]{1997Forbes, 2006Harris},
are significantly higher than $-3$. 
This indicates that, in scenarios where globular clusters are formed in dwarf,
atomic cooling galaxies \citep[][]{2016Kimm},
most of the metals ought to originate from previous generation of stars formed in either other and/or progenitor atomic cooling 
halos prior to forming globular clusters at the centers of these dwarf galaxies. 
Given the much reduced star formation hence metallicity 
in minihalos, it would seem conceivable that Pop III stars formed in atomic cooling halos 
may make a significant contribution to the pre-enrichment (of, say, Fe) 
of the gas forming the first-generation stars in globular clusters.

\medskip
I thank Zoltan Haiman, Kohei Inayoshi, Taysum Kimm, John Wise 
and Jemma Wolcott-Green for useful discussion and communications,
Paola Marigo for stellar track data,
Alexander Heger, Berhnard Mueller and Ondrej Pejcha for educational discussion on supernova related issues,
and Umberto Maio for sharing low temperature cooling data files.
This work is supported in part by grants NNX12AF91G and AST15-15389.

%%%%%%%%%%%%%%%%%%%%%%%%%%%%%%%%%%%%%%%%%%%%%%%%%%

%%%%%%%%%%%%%%%%%%%% REFERENCES %%%%%%%%%%%%%%%%%%

% The best way to enter references is to use BibTeX:

%\bibliographystyle{apj}
%\bibliographystyle{mnras}
%\bibliography{astro} % if your bibtex file is called example.bib

%Must run the following (change typeset option from LaTeX to BibTex)
% latex <file>
% bibtex <file>
% latex <file>
% latex <file>
%%%%%%%%%%%%%%%%%%%%%%%%%%%%%%%%%%%%%%%%%%%%%%%%%%

%%%%%%%%%%%%%%%%% APPENDICES %%%%%%%%%%%%%%%%%%%%%

%\appendix

%\section{Some extra material}

%%%%%%%%%%%%%%%%%%%%%%%%%%%%%%%%%%%%%%%%%%%%%%%%%%

% Don't change these lines
%\bsp   % typesetting comment
\label{lastpage}
\end{document}